\documentclass[epj,nopacs]{svjour}
\usepackage{graphics}
\usepackage[english]{babel}
\usepackage[dvips]{graphicx}
\usepackage{verbatim}
\usepackage{makeidx}
\usepackage{color}
\usepackage{amsfonts}
\usepackage{amsmath}

\newcommand{\sqrtsNN}{\sqrt{s_{\scriptscriptstyle{{\rm NN}}}}}
\newcommand{\av}[1]{\left\langle #1 \right\rangle}

\newcommand{\mev}{\mathrm{MeV}}
\newcommand{\gev}{\mathrm{GeV}}
\newcommand{\tev}{\mathrm{TeV}}
\newcommand{\fm}{\mathrm{fm}}

\newcommand{\mum}{\mathrm{\mu m}}

\newcommand{\PbPb}{\mbox{Pb--Pb}}

\newcommand{\NN}{\mbox{nucleon--nucleon}}

\newcommand{\RAA}{R_{\rm AA}}
\newcommand{\RDh}{R_{{\rm D}/h}}
\newcommand{\pt}{p_{\rm t}}
\renewcommand{\d}{{\rm d}}

\newcommand{\ccbar}{\mbox{$\mathrm {c\overline{c}}$}}

\newcommand{\Dz}{\mbox{$\mathrm {D^0}$}}
\newcommand{\DtoKpi}{\mbox{${\rm D^0\to K^-\pi^+}$}}

\begin{document}
\title{Perspectives for the study of charm in-medium quenching
at the LHC with ALICE}
\author{Andrea Dainese}
%
\institute{Universit\`a degli Studi di Padova, 
        via Marzolo 8, 35131 Padova, Italy; e-mail: andrea.dainese@pd.infn.it}
\date{Received: 4 December 2003 / Published online: 3 March 2004}
%
\abstract{
Charm mesons produced in nucleus--nucleus collisions are expected to be less 
attenuated (quenched) by the medium than hadrons containing only light quarks,
since radiative energy loss of heavy quarks should be reduced
by the `dead-cone' effect.
We start from a published energy-loss model to derive the quenching for 
D mesons at the LHC, 
introducing an approximation of the dead-cone effect and employing 
a Glauber-based description of the geometry of central Pb--Pb collisions 
to estimate the in-medium path lengths of c quarks.
We show that the exclusive 
reconstruction of $\DtoKpi$ decays in ALICE allows to measure
the nuclear modification factor of the D mesons transverse momentum  
distribution and the D/$charged~hadrons$ ratio and, thus, to investigate the 
energy loss of c quarks.
} 
\maketitle

\section{Introduction}
\label{intro}

The ALICE experiment~\cite{tpalice} at the LHC 
will study nucleus--nucleus (AA) collisions at a centre-of-mass energy
$\sqrtsNN=5.5~\tev$ (for \PbPb) per nucleon--nucleon (NN) pair in order to 
investigate the properties of QCD matter at energy densities of 
up to several hundred times the density of atomic nuclei. In these conditions
a deconfined state of quarks and gluons is expected to be formed.

Hard partons and heavy quarks, abundantly produced at LHC energies in 
initial hard-scattering processes, are sensitive probes of the medium 
formed in the collision as they may lose energy by gluon bremsstrahlung 
while propagating through the 
medium~\cite{gyulassywang,bdmps,zakharov,wiedemann}. 
The attenuation (quenching) of leading hadrons and jets observed at 
RHIC~\cite{phenixRAA1,phenixRAA2,starRAA,starAzi} 
is thought to be due to such a mechanism.
The large masses of the charm and beauty quarks make them qualitatively 
different probes, since, on well-established QCD grounds, in-medium energy loss
of massive partons is expected to be significantly smaller than that 
of `massless' partons (light quarks and gluons)~\cite{dokshitzerkharzeev}. 
Therefore, a comparative study of the attenuation of massless and 
massive probes is a promising tool to test the consistency 
of the interpretation 
of quenching effects as energy loss in a deconfined medium and to further 
investigate the properties (density) of such medium.

In the first part of this paper, we shortly summarize a widely 
used model of parton energy loss and we discuss how we apply it in our 
simulation. In the second part, we show that the exclusive reconstruction 
of $\DtoKpi$ decays with ALICE allows to carry out the mentioned comparative 
quenching studies by measuring:
\begin{itemize} 
\item the {\it nuclear modification factor} of 
D mesons as a function of transverse momentum ($\pt$)
\begin{equation}
\label{eq:raa}
  R_{\rm AA}(\pt)\equiv
    \frac{{\rm d}N_{\rm AA}/{\rm d}\pt/{\rm binary~NN~collision}}
       {{\rm d}N_{\rm pp}/{\rm d}\pt},
\end{equation}
which would be equal to 
1 if the AA collision was a mere superposition of independent 
NN collisions, without nuclear or medium effects; in central Au--Au collisions
at RHIC ($\sqrtsNN=200~\gev$) $\RAA\simeq 0.2$ for both $\pi^0$ and 
charged hadrons in the range 
$4<\pt<8~\gev/c$~\cite{phenixRAA1,phenixRAA2,starRAA};
\item the ratio of the nuclear modification factors of D mesons and
of charged (non-charm) hadrons, as a function of $\pt$:
\begin{equation}
\label{eq:RDh}
   R_{{\rm D}/h}(\pt)\equiv R_{\rm AA}^{\rm D}(\pt)\Big/R_{\rm AA}^h(\pt);
\end{equation}
hereafter, this quantity is called D/$charged~hadrons$ (D/$h$) ratio. 
\end{itemize}

\section{Parton energy loss and the dead-cone effect for heavy quarks}
\label{energyloss}

In this work we use the quenching probabilities (or weights) calculated in 
Ref.~\cite{carlosurs} in the framework of the `BDMPS' 
(Baier-Dokshitzer-Mueller-Peign\'e-Schiff) 
formalism~\cite{bdmps} summarized in the following.
The energy loss obtained with the quenching weights is presented in 
Section~\ref{simulation}.

An energetic parton produced in a hard collision radiates a
gluon with a probability proportional to its path length $L$
in the dense medium. Then,
the radiated gluon undergoes multiple scatterings in the medium, in 
a Brown\-ian-like motion with mean free path $\lambda$ which decreases 
as the density of the medium increases. The number of scatterings of the 
radiated gluon is also proportional to $L$. Therefore, the average energy loss
of the parton is proportional to $L^2$.

The scale of the energy loss is set by the `maximum' energy of the radiated
gluons, which depends on $L$ and on the properties of the 
medium:
\begin{equation}
 \omega_c = \hat{q}\,L^2/2,
\end{equation}
where $\hat{q}$ is the transport coefficient of the medium, defined
as the average transverse momentum squared transferred to the projectile 
per unit path length, 
$\hat{q} = \av{q_{\rm t}^2}_{\rm medium}\big/\lambda$~\cite{carlosurs}.

In the case of a static medium, the distribution of the energy $\omega$ of the
radiated gluons (for $\omega\ll\omega_c$) is of the form:
\begin{equation} 
\label{eq:wdIdw}
\omega\frac{{\rm d}I}{{\rm d}\omega}\simeq \frac{2\,\alpha_s\,C_R}{\pi}\sqrt{\frac{\omega_c}{2\omega}},
\end{equation}
where $C_R$ is the QCD coupling factor (Casimir factor), equal 
to 4/3 for quark--gluon coupling and to 3 for gluon--gluon coupling.
The integral of the energy distribution up to $\omega_c$ estimates the 
average energy loss of the parton:
\begin{equation}
\label{eq:avdE}
\av{\Delta E} = \int_0^{\omega_c} \omega \frac{{\rm d}I}{{\rm d}\omega}{\rm d}\omega
\propto \alpha_s\,C_R\,\omega_c \propto \alpha_s\,C_R\,\hat{q}\,L^2.
\end{equation}
The average energy loss is:
proportional to $\alpha_s\,C_R$ and, thus, larger by a factor 
$9/4=2.25$ for gluons than for quarks; 
proportional to the transport coefficient of the medium;
proportional to $L^2$; 
independent of the parton initial energy $E$.  
It is a general feature of all parton energy loss 
calculations~\cite{gyulassywang,bdmps,zakharov,wiedemann,carlosurs,gyulassy2}
that the energy distribution~(\ref{eq:wdIdw}) does not depend on $E$.
Depending on how the kinematic bounds are taken into account, the resulting
$\Delta E$ is $E$-independent~\cite{bdmps} or depends logarithmically on 
$E$~\cite{gyulassy2}.
However, there is always a stronger 
intrinsic dependence of the radiated energy on 
the initial energy, determined by the fact that the former cannot be larger 
than the latter, $\Delta E\leq E$.
Within the above toy-model derivation which agrees with the main features of 
the BDMPS formalism, this kinematic
constraint could be included by truncating the gluon energy distribution 
$\omega\,\d I/\d\omega$ at $\min(\omega_c,E)$ rather than at its natural 
upper limit $\omega_c$. This would give, from (\ref{eq:wdIdw}) and 
(\ref{eq:avdE}), $\av{\Delta E}\propto \alpha_s\,C_R\,\sqrt{\omega_c}\,\sqrt{\min(\omega_c,E)}$. For $E<\omega_c$, we have 
$\av{\Delta E}\propto\sqrt{\hat{q}}\,\sqrt{E}\,L$: the kinematic constraint
turns the $L$-dependence from quadratic to linear. 
As we shall discuss in Section~\ref{simulation},
the kinematic constraint can be equivalently interpreted as a reduction of 
the effective path length in the medium, i.e. of the length along which 
the energy of the parton is larger than zero and gluons can be radiated. 

The transport coefficient is proportional to the density of the 
scattering centres and to the typical momentum transfer in the 
gluon scattering off these centres.
A review of the estimates for the value of the transport coefficient
in media of different densities can be found in Ref.~\cite{YRjets}: 
the estimate is $\hat{q}_{\rm cold} \simeq 0.05~\gev^2/\fm$ for cold nuclear
matter and up to $0.5~\gev^2/\fm$ for a hadron gas; 
for a QGP formed at the LHC with energy density 
$\epsilon\sim 50$--$100~\gev/\fm^3$, $\hat{q}$ is expected to be 
of $\simeq 5$--$10~\gev^2/\fm$. 

The medium-induced energy loss of heavy quarks was first studied in 
Refs.~\cite{mustafa,linvogt}. 
Later, in Ref.~\cite{dokshitzerkharzeev} it was argued that for heavy quarks, 
because of their large mass, the radiative energy loss should be 
lower than for light quarks. The predicted consequence of this effect was
an enhancement of the ratio of D mesons to pions (or hadrons in general) 
at moderately-large ($5$--$10~\gev/c$) transverse momenta, with respect to that
observed in the absence of energy loss. 

Heavy quarks with momenta up to $40$--$50~\gev/c$ propagate with 
a velocity which is significantly smaller than the velocity of light. 
As a consequence, in the vacuum, 
gluon radiation at angles $\Theta$ smaller than the ratio of their mass to 
their energy $\Theta_0=m/E$ is suppressed by destructive quantum 
interference~\cite{dokshitzerdeadcone}. 
The relatively depopulated cone around the heavy-quark direction with 
$\Theta<\Theta_0$ is called `dead cone'.

In Ref.~\cite{dokshitzerkharzeev} the dead-cone effect is assumed to 
characterize also in-medium gluon radiation and 
the energy distribution of the radiated gluons (\ref{eq:wdIdw}),
for heavy quarks, is estimated to be suppressed by the factor:
\begin{equation}
  \label{eq:FHL}
  \begin{array}{rcl}
  \displaystyle{\frac{{\rm d}I}{{\rm d}\omega}\bigg|_{\rm Heavy}\bigg/
  \frac{{\rm d}I}{{\rm d}\omega}\bigg|_{\rm Light}} & = &
  \displaystyle{\left[1+\frac{\Theta_0^2}{\Theta^2}\right]^{-2}}\\
  & = &
  \displaystyle{\left[1+\left(\frac{m}{E}\right)^2\sqrt{\frac{\omega^3}{\hat{q}}}\right]^{-2}}\\
  & \equiv & \displaystyle{F_{\rm H/L}}(m,E,\omega,\hat{q})\\
\end{array}
\end{equation} 
where the expression for the characteristic gluon emission 
angle~\cite{dokshitzerkharzeev} 
$\Theta\simeq (\hat{q}/\omega^3)^{1/4}$ has been used.
The heavy-to-light suppression factor $F_{\rm H/L}$ in (\ref{eq:FHL}) 
increases (less suppression) as the 
heavy-quark energy $E$ increases (the mass becomes negligible) 
and it decreases at large $\omega$, indicating that the high-energy 
part of the 
gluon radiation spectrum is drastically suppressed by the dead-cone effect.

The first (indirect) measurements of charm production, 
in the semi-elec\-tronic decay 
channel, in Au--Au collisions at RHIC~\cite{phenixcharm} 
indicate (within the still large 
experimental errors) that there may be no attenuation of the D meson 
$\pt$ distribution, contrarily to what observed for pions and charged hadrons. 
This result supports the scenario proposed in Ref.~\cite{dokshitzerkharzeev}
and it has stimulated considerable theoretical interest in the 
subject~\cite{gyulassyheavy,wangheavy}. Our approach, described in the next 
section, is to implement in the energy-loss simulation an algorithm to account 
for the dead-cone effect.

\section{Simulation of energy loss}
\label{simulation}

The quenching weight~\cite{bdmsJHEP,carlosurs} is defined as the probability 
that a hard parton radiates an energy $\Delta E$ due to scattering in 
spatially-extended QCD matter. In Ref.~\cite{carlosurs}, the weights are 
calculated on the basis of the BDMPS formalism, taking into account both  
the finite in-medium path length $L$ and the dynamic expansion of 
the medium after the nucleus--nucleus collision.
The input parameters for the calculation are the 
length $L$, the transport coefficient $\hat{q}$ and the parton species
(light quark or gluon).

The distribution of the in-medium path length in the plane transverse 
to the beam line\footnote{Partons produced at central rapidities 
propagate in the transverse plane.} for central
\PbPb~collisions (impact parameter $b<3.5~\fm$, corresponding to the 5\% most 
central collisions) is calculated in the framework
of the Glauber model for the collision geometry~\cite{glauber}. 
For a given impact parameter,
hard-parton production points are sampled according 
to the density $\rho_{\rm coll}(x,y)$ of binary nucleon--nucleon collisions in 
the transverse plane and their 
azimuthal propagation directions are sampled uniformly. 
For a parton with production point $(x_0,y_0)$ 
and azimuthal direction  $(u_x,u_y)$, 
the path length is defined as:
\begin{equation}
\label{eq:ell}
L = \frac{\int_0^\infty {\rm d}l\,l\,\rho_{\rm coll}(x_0+l\,u_x,y_0+l\,u_y)}
    {0.5\,\int_0^\infty {\rm d}l\,   \rho_{\rm coll}(x_0+l\,u_x,y_0+l\,u_y)}.
\end{equation} 

\begin{figure}[!t]
  \begin{center}
    \includegraphics[width=.48\textwidth]{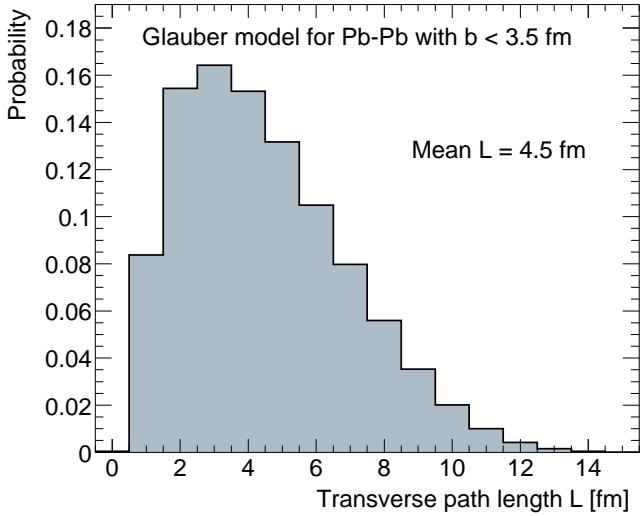}
    \caption{Distribution of the path lengths in the transverse plane 
             for partons produced in \PbPb~collisions with impact parameter 
             $b<3.5~\fm$.} 
    \label{fig:ell}
  \end{center}
\end{figure}

Many sampling iterations are performed varying the impact 
parameter $b$ from $0.25~\fm$ to $3.25~\fm$ in steps of $0.5~\fm$. 
The obtained distributions are given a weight $b$, 
since we verified that 
d$\sigma^{\rm hard}/$d$b\propto b$ for $b<3.5~\fm$, and summed. 
The result is shown in Fig.~\ref{fig:ell}. 
The average length is $4.5~\fm$, corresponding to 
about 70\% of the radius of a Pb nucleus and 
the distribution is significantly shifted towards low values of $L$
because a large fraction of the partons are produced in the periphery
of the superposition region of the two nuclei (`corona' effect).

The definition of $L$ in (\ref{eq:ell}) 
is exact in the case of a cylindrical collision density 
profile (i.e. $\rho_{\rm coll}=\rho_0\,\Theta(R-\sqrt{x^2+y^2})$ for a 
cylinder with radius $R$). It is approximated for a realistic profile
(derived from a Wood-Saxon nuclear density profile). The inclusion of nuclear
geometry effects is not straight-forward in the scheme of the 
quenching weights, where the transport coefficient and the path length 
are considered as two distinct input parameters. Indeed, the lower 
medium density in the `corona' could be modeled with a reduction of the 
transport coefficient or of the path length. However, since the energy 
loss is determined by the quantities $\omega_c\propto\hat{q}\,L^2$ and
$\omega_c\,L\propto\hat{q}\,L^3$~\cite{carlosurs}, the optimal solution 
would be to include the geometry profile directly in the calculation of 
these two parameters, defining them as line integrals in $\d l$ similar to 
the integrals that appear in (\ref{eq:ell}). 
For this work we adopt (\ref{eq:ell}) as a practical definition of length 
in the medium, which allows us to employ a distribution of $L$ rather than
a constant value; in the following we show that this point is quite
relevant.  
A more refined treatment of nuclear geometry is discussed in 
Ref.~\cite{drees}, where, however, energy loss is modeled as a simple
exponential absorption that cannot be directly related to the medium 
properties. 

\begin{figure*}[!t]
  \begin{center}
    \includegraphics[width=.48\textwidth]{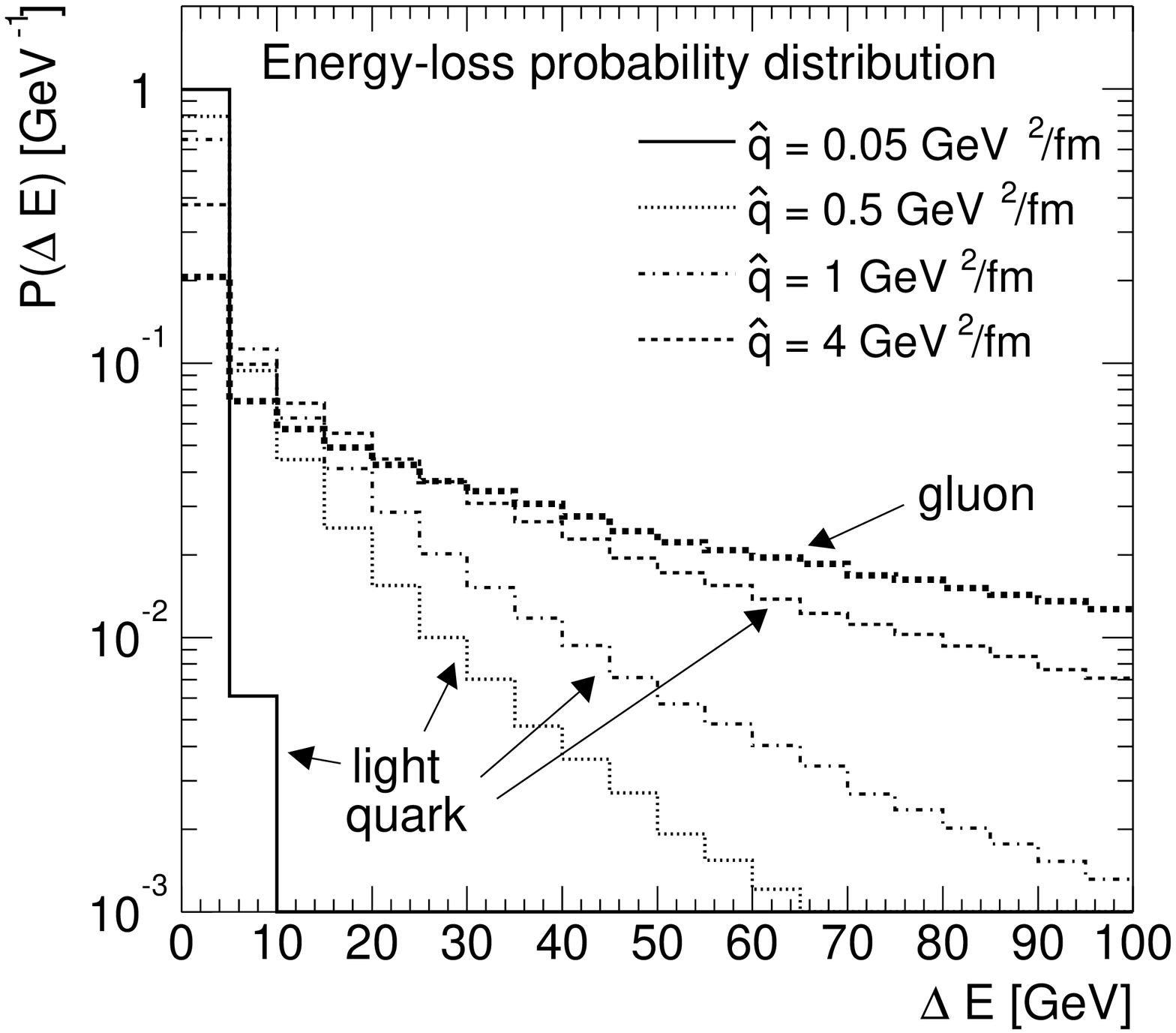}
    \includegraphics[width=.48\textwidth]{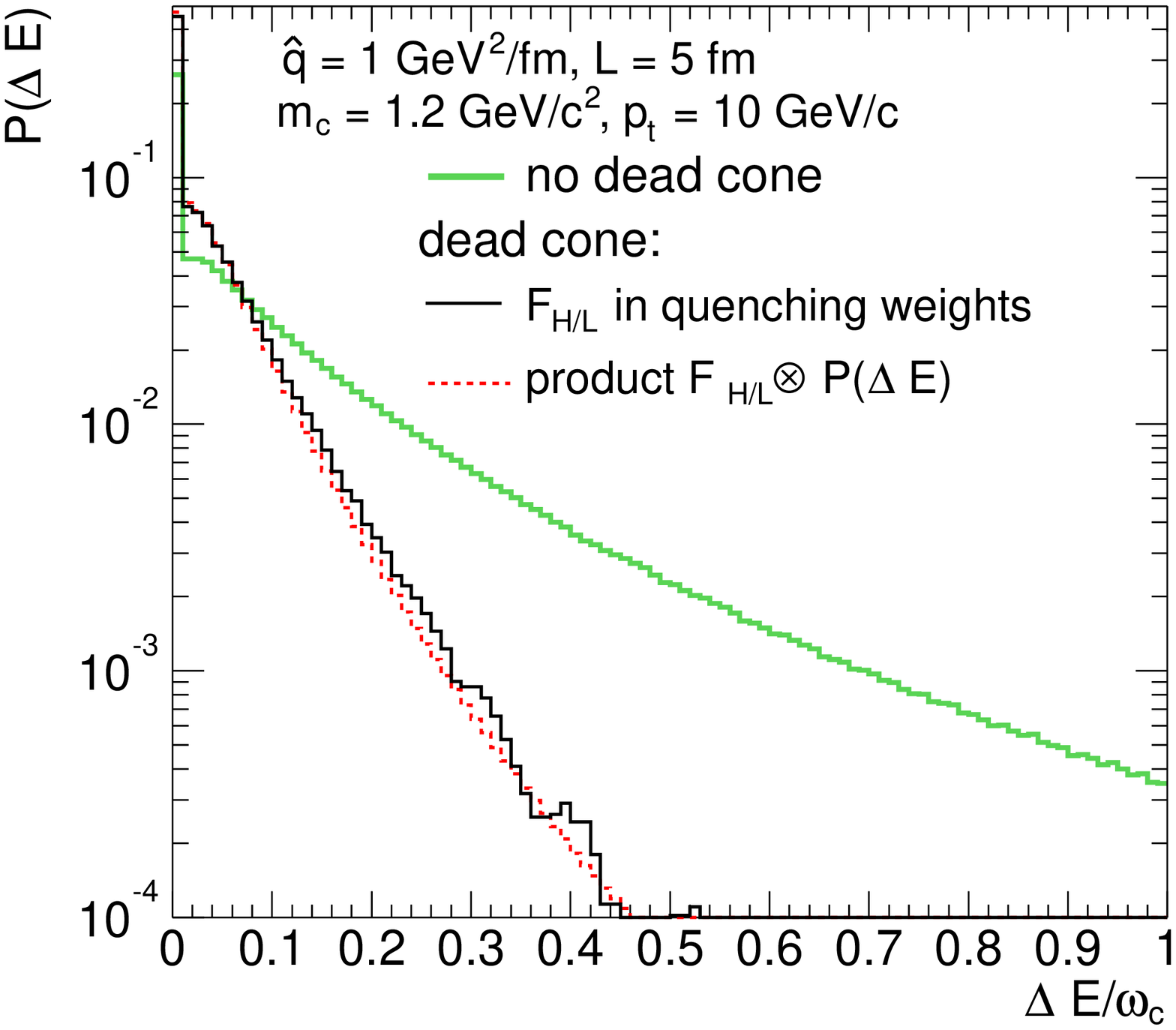}
    \caption{Left-hand panel: 
             energy-loss probability distribution, $P(\Delta E)$, for 
             different values of $\hat{q}$, for light quarks and for gluons
             (thicker dashed line, only for $4~\gev^2/\fm$). Right-hand panel:
             $P(\Delta E)$ for c quarks ($m_{\rm c}=1.2~\gev/c^2$, 
             $\pt=10~\gev/c$) 
             without dead cone and with two different 
             implementations of the dead-cone effect (see text).} 
    \label{fig:PDeltaE}
  \end{center}
\end{figure*}

For a given value of the transport coefficient $\hat{q}$ and a given parton 
species, we use the numerical routine 
provided in Ref.~\cite{carlosurs} to calculate the energy-loss probability 
distribution $P(\Delta E;L)$ for the integer values of $L$ up to 
$15~\fm$. Then, these 15 distributions are 
weighted according to the path-length probability in Fig.~\ref{fig:ell}
and summed up to obtain a global energy-loss probability distribution 
$P(\Delta E)$. The energy loss to be used for the quenching simulation
can be directly sampled from the $P(\Delta E)$ distribution 
corresponding to the chosen $\hat{q}$ and to the correct parton species.
Figure~\ref{fig:PDeltaE} (left) reports $P(\Delta E)$ for light quarks 
and for gluons, as obtained with different values of the transport
coefficient; in the figure, the `peak' at $\Delta E=0$ represents the 
probability to have no medium-induced gluon radiation. 

The predicted lower energy loss for charm quarks is accounted for  
by multiplying the $P(\Delta E)$ distribution for light 
quarks by the dead-cone suppression factor 
$F_{\rm H/L}$ given in (\ref{eq:FHL}), with $\omega=\Delta E$.
Since $F_{\rm H/L}$ depends on the heavy-quark energy $E$, the
multiplication 
has to be done for each c quark or, more conveniently, in bins of 
the quark energy.
It was verified that this multiplication is equivalent to 
recalculating the quenching weights with the gluon energy distribution 
for heavy quarks modified according to (\ref{eq:FHL}).  The results 
on the energy-loss probability distribution obtained with the two methods, 
`$F_{\rm H/L}\otimes P(\Delta E)$ product' and `$F_{\rm H/L}$ in the quenching
weights', were compared for
$\hat{q}=1~\gev^2/\fm$, $L=5~\fm$, $m_{\rm c}=1.2~\gev/c^2$ and 
$\pt=10,~20,~30~\gev/c$~\cite{carlosurspriv}. 
The comparison for $\pt=10~\gev/c$, 
shown in Fig.~\ref{fig:PDeltaE} (right), is quite satisfactory.
A similar agreement is found for $\pt=20$ and $30~\gev/c$. 

\begin{figure*}[!t]
  \begin{center}
    \includegraphics[width=.95\textwidth]{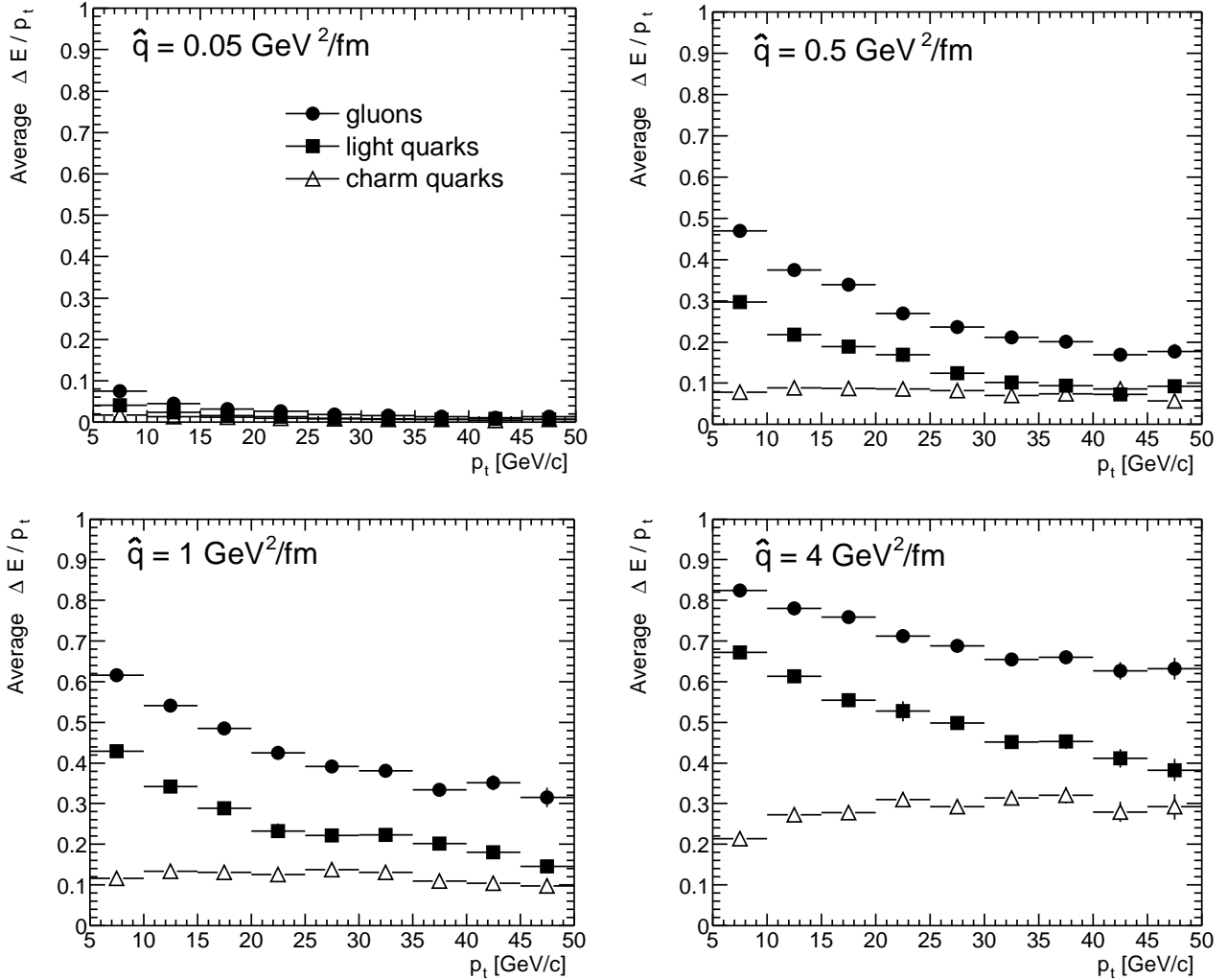}
    \caption{Average relative energy loss as function of the transverse 
             momentum for gluons, light (massless) quarks and charm 
             quarks ($m_{\rm c}=1.2~\gev/c^2$).}
    \label{fig:avreldE}
  \end{center}
\end{figure*}

Figure~\ref{fig:avreldE} shows the average relative energy 
loss as a function of the transverse momentum for gluons, light quarks and 
charm quarks ($m_{\rm c}=1.2~\gev/c^2$). 
Different values of the transport coefficient $\hat{q}$ are considered 
in the various panels and the 
described dead-cone correction (energy-dependent
$F_{\rm H/L}\otimes P(\Delta E)$ product) is used for charm quarks. 
For each value of the parton $\pt$ many values of $\Delta E$ are sampled 
according to the energy-loss probability distribution and the kinematic 
constraint $\Delta E\leq E$, discussed in Section~\ref{energyloss}, 
is applied as follows: if $\Delta E>\pt$, then $\Delta E=\pt$.
The ratio of the relative energy losses for gluons and for light quarks is 
compatible with the ratio of their Casimir factors (2.25) at high $\pt$, where
the kinematic constraint is not relevant. The ratio of the relative energy 
losses for light quarks and c quarks increases with increasing $\hat{q}$, 
particularly in the high-$\pt$ region, showing that the dead-cone effect is
medium-dependent. 
With $\hat{q}=4~\gev^2/\fm$, our estimated transport coefficient for the 
LHC (see next paragraph), the average relative energy loss is 
$\approx (85-0.6\,\pt(\gev/c))\%$ for gluons, 
$\approx (75-0.8\,\pt(\gev/c))\%$ for light quarks
and $\approx 25$--$30\%$ for c quarks.

Remarkably, for charm quarks we find that, for given $\hat{q}$,
the average relative energy loss is approximately independent of the quark 
energy, $\av{\Delta E/E}\approx const.$, 
(see Fig.~\ref{fig:avreldE}), while for massless partons the average
relative energy loss is clearly decreasing as the parton energy increases
(the BDMPS average relative energy loss for massless partons would be 
$\av{\Delta E/E}\propto 1/E$; this dependence is, then, 
weakened by the kinematic constraint). On the basis of this 
observation we expect that not only the magnitude but also 
the $\pt$-dependence of the nuclear modification factor $\RAA$ of D mesons 
can be significantly affected by the dead cone.

Before describing the estimation of the transport coefficient to be used 
in the simulation, we shortly comment on the implementation of the kinematic
constraint. Taking, as we do, $\Delta E=\pt$ when the sampled $\Delta E$
is larger than $\pt$ is equivalent to truncating the energy-loss probability 
distribution $P(\Delta E)$ at $\Delta E=\pt$ and adding the $\delta$-function
$\delta(\Delta E-\pt)\int_{\pt}^\infty\d\mathcal{E}\,P(\mathcal{E})$
to it. The total integral of $P$ is, in this way, maintained equal to 1.
The dependence of the final result on the kinematic constraint was 
argued~\cite{wiedemann2,carlosurs} to illustrate the theoretical uncertainties.
These can be sizeable (more than a factor 2) for low $\pt$ and sufficiently
large $\hat{q}$ and $L$ in the present study. They should be improved
in a refined calculation but we did not try to quantify them in 
detail\footnote{These topics are considered also in Ref.~\cite{ehsw} of 
which the author became
aware while finalizing the present work. Very useful discussions with the 
authors of Ref.~\cite{ehsw} are acknowledged.}. 
~\\

For the estimation of the transport coefficient $\hat{q}$ 
we require for central nucleus--nucleus 
collisions at the LHC a leading-particle quenching
at least of the same magnitude as that observed at RHIC.
We, therefore, derive the nuclear modification factor $\RAA$ for charged 
hadrons produced at the LHC and we choose the transport coefficient in 
order to obtain $\RAA\simeq 0.2$--$0.3$ in the range $5<\pt<10~\gev/c$
(for RHIC results see e.g. Refs.~\cite{phenixRAA1,phenixRAA2,starRAA}). 

\begin{figure*}[!t]
  \begin{center}
    \includegraphics[width=.95\textwidth]{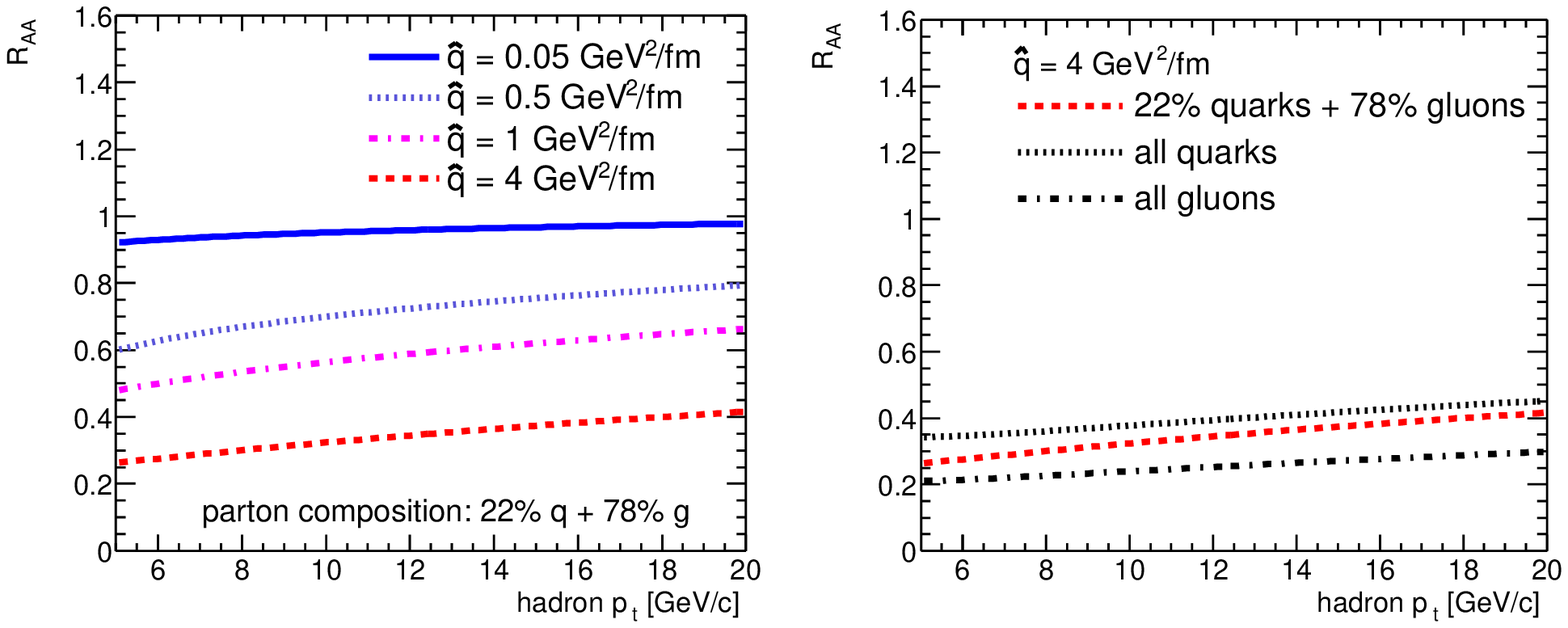}
    \caption{Nuclear modification factor for charged hadrons 
             for different values of $\hat{q}$ (left) and comparison 
             of quark and gluon quenching for $\hat{q}=4~\gev^2/\fm$ (right).} 
    \label{fig:RAA-hadrons}
\vglue0.5cm
    \includegraphics[width=.95\textwidth]{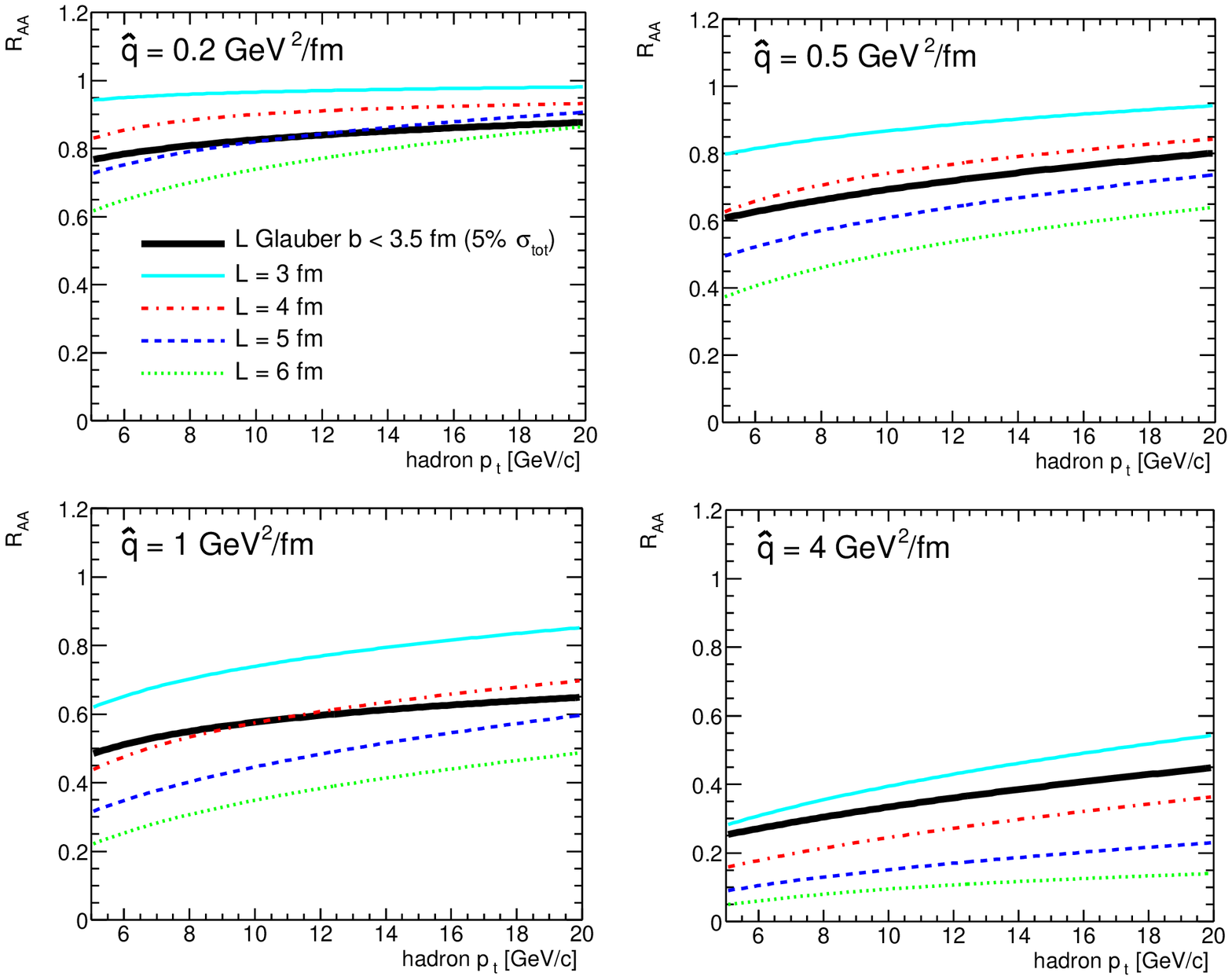}
    \caption{Nuclear modification factor for charged hadrons 
             for different values of $\hat{q}$ as obtained using the complete
             Glauber-based distribution (from Fig.~\ref{fig:ell})
             or constant values of $L$.} 
    \label{fig:RAA-hadrons-cmpL}
  \end{center}
\end{figure*}

The transverse momentum distributions, for $\pt>5~\gev/c$, of charged hadrons 
are generated by means of the chain:
\begin{enumerate}
\item generation of a parton, quark or gluon, with $\pt>5~\gev/c$, 
      using PYTHIA~\cite{pythia} proton--proton 
      with $\sqrt{s}=5.5~\tev$ and CTEQ 4L 
      parton distribution functions~\cite{cteq4}; 
      with these parameters, the parton 
      composition given by PYTHIA is 78\% gluons and 22\% quarks (average over
      $\pt>5~\gev/c$);
\item sampling of an energy loss $\Delta E$ according to $P(\Delta E)$ 
      and calculation of the new 
      transverse momentum of the parton, $\pt'=\pt-\Delta E$ (if 
      $\Delta E>\pt$, $\Delta E=\pt$ and $\pt'=0$);
\item (independent) fragmentation of the parton to a hadron using the 
      leading-order Kniehl-Kramer-P\"otter (KKP) fragmentation 
      functions~\cite{kkp}. 
\end{enumerate}
Quenched and unquenched $\pt$ distributions are obtained including or 
excluding the second step of the chain.
Figure~\ref{fig:RAA-hadrons} shows $\RAA$ for 
hadrons, calculated as the ratio of the $\pt$ distribution with quenching 
to the $\pt$ distribution without quenching. Different values of $\hat{q}$
are considered in the left-hand panel of the figure: a value as large as 
$4~\gev^2/\fm$ is necessary to have $\RAA\simeq 0.25$--$0.3$ in 
$5<\pt<10~\gev/c$. In the right-hand panel, for $\hat{q}=4~\gev^2/\fm$, we 
compare the results obtained considering all partons as gluons or all 
partons as quarks, in order to remark and quantify the larger quenching 
of gluons with respect to quarks.

Since the transport coefficient determines the size of the energy-loss
effect, we shortly discuss the choice of $\hat{q}=4~\gev^2/\fm$.
This value corresponds, according to the estimates reported in 
Ref.~\cite{YRjets}, to an energy density 
$\epsilon\simeq 40$--$50~\gev/\fm^3$, which is about a factor 2 lower 
than the maximum energy density expected for central \PbPb~collisions at the
LHC. The value is, therefore, reasonable.

Using the same quenching weights, the suppression observed at RHIC is 
reproduced in Ref.~\cite{carlosurs} with 
the much lower value $\hat{q}=0.75~\gev^2/\fm$. However, 
the constant length $L=6~\fm$ is there used, 
rather than a realistic distribution 
of lengths, thus obtaining a significantly stronger quenching. This simple
approximation captures the main features of $\RAA$ for 5--10\% central 
collisions, which depend on the combination $\hat{q}\,L^3$ 
($\propto\omega_c\,L$) more than  
on $\hat{q}$ and $L$ separately~\cite{carlosurs}. In addition, transverse
momentum distributions are steeper at RHIC energy
than at LHC energy and, consequently, to obtain the same $\RAA$ suppression 
in the two cases one needs a larger $\hat{q}$ (or $\av{\Delta E}$) 
at the LHC than at RHIC. 

Because of the kinematic constraint $\Delta E\leq E$,
the use of a constant length of the order of the nuclear radius or even 
the use of the average length from a detailed distribution can produce quite 
different results with respect to those obtained taking into account 
the complete distribution. This is demonstrated in 
Fig.~\ref{fig:RAA-hadrons-cmpL}: for $\hat{q}=0.5$--$1~\gev^2/\fm$, 
$L=6~\fm$ gives almost a factor 2 difference in $\RAA$ at $\pt\sim 10~\gev/c$
and the complete $L$ distribution is equivalent to a constant length which 
decreases as $\hat{q}$ increases, 5, 4.5, 4, 3.5~fm for 
$\hat{q}=0.2,~0.5,~1,~4~\gev^{2}/\fm$. This behaviour is clearly due to 
an upper `cut-off' of the length distribution: large lengths correspond to 
very high values of $\Delta E$, but, since $\Delta E$ cannot be higher than 
the initial parton energy $E$, large lengths are not `fully exploited'; 
this corresponds to a cut-off; e.g. for many partons of moderate energy a 
length of $8~\fm$ is equivalent to a length of $4~\fm$, because after
propagating for $4~\fm$ they have lost all their initial energy. 
As a consequence,
the length distribution corresponds to an average `effective' length lower than
its arithmetic average. The cut-off moves towards lower lengths as
$\hat{q}$ increases and, thus, the average effective length decreases.
For $\hat{q}=4~\gev^2/\fm$, the average effective energy loss is 
approximately: $\av{\Delta E_{\rm eff}}\approx\av{\Delta E}\times\av{L_{\rm eff}}^2/\av{L}^2=\av{\Delta E}\times (3.5/4.5)^2\approx 0.6\,\av{\Delta E}$.

Another important observation revealed by Fig.~\ref{fig:RAA-hadrons-cmpL} 
is the fact that the use of the complete $L$ distribution reduces 
the increase of $\RAA$ with $\pt$ (at RHIC, $\RAA$ is found to be independent
of $\pt$ in the range $4$--$10~\gev/c$~\cite{phenixRAA2}).
This happens because higher-energy partons can exploit the large-$L$ tail more
than lower-energy partons and, consequently, for them the cut-off is shifted 
towards larger lengths.

Charm quarks are generated using PYTHIA, tuned in order to reproduce the 
single-inclusive c (and $\overline{\rm c}$) $\pt$ distribution predicted 
by the pQCD program 
HVQMNR~\cite{hvqmnr} with $m_{\rm c}=1.2~\gev/c^2$ and factorization and 
renormalization scales
$\mu_F=\mu_R=2\,m_{\rm t}\equiv 2\,\sqrt{m_{\rm c}^2+\pt^2}$
(the details on this tuning can be found in Ref.~\cite{noteHVQprod}).
In HVQMNR we use the CTEQ 5M parton distribution functions~\cite{cteq5} 
including, for \PbPb, the nuclear
shadowing effect by means of the EKS98 parameterization~\cite{EKS}
and the parton intrinsic transverse momentum broadening as reported
in Ref.~\cite{noteHVQprod}. With these parameters, the $\ccbar$ production 
yields in pp collisions at $\sqrt{s}=14~\tev$ and in 
central (5\%) \PbPb~collisions at $\sqrtsNN=5.5~\tev$ are 
$N^{\scriptstyle{\rm c\overline{c}}}_{\rm pp}=0.16$ and
$N^{\scriptstyle{\rm c\overline{c}}}_{\rm Pb-Pb}=115$, respectively.
The yield given for \PbPb~already includes a 65\% reduction due to shadowing.

Energy loss for charm quarks is simulated following a slightly different 
procedure with respect to that for light quarks and gluons. Since
the total number of $\ccbar$ pairs per event has to be conserved, in the 
cases where the sampled $\Delta E$ is larger than $\pt$, we assume the c 
quark to be thermalized in the medium and we give it a transverse momentum 
according to the distribution 
d$N/$d$m_{\rm t}\propto m_{\rm t}\,\exp(-m_{\rm t}/T)$, as 
suggested in Ref.~\cite{linvogt}. 
We use $T=300~\mev$ as the thermalization temperature for c quarks. 
The other difference with respect to the case of massless partons 
is that we use the standard string model in PYTHIA for the c-quark 
fragmentation (more details can be found in Ref.~\cite{thesis}).

\section{Charm reconstruction with ALICE}
\label{D0reco}

\begin{figure*}[!t]
  \begin{center}
    \includegraphics[width=.8\textwidth]{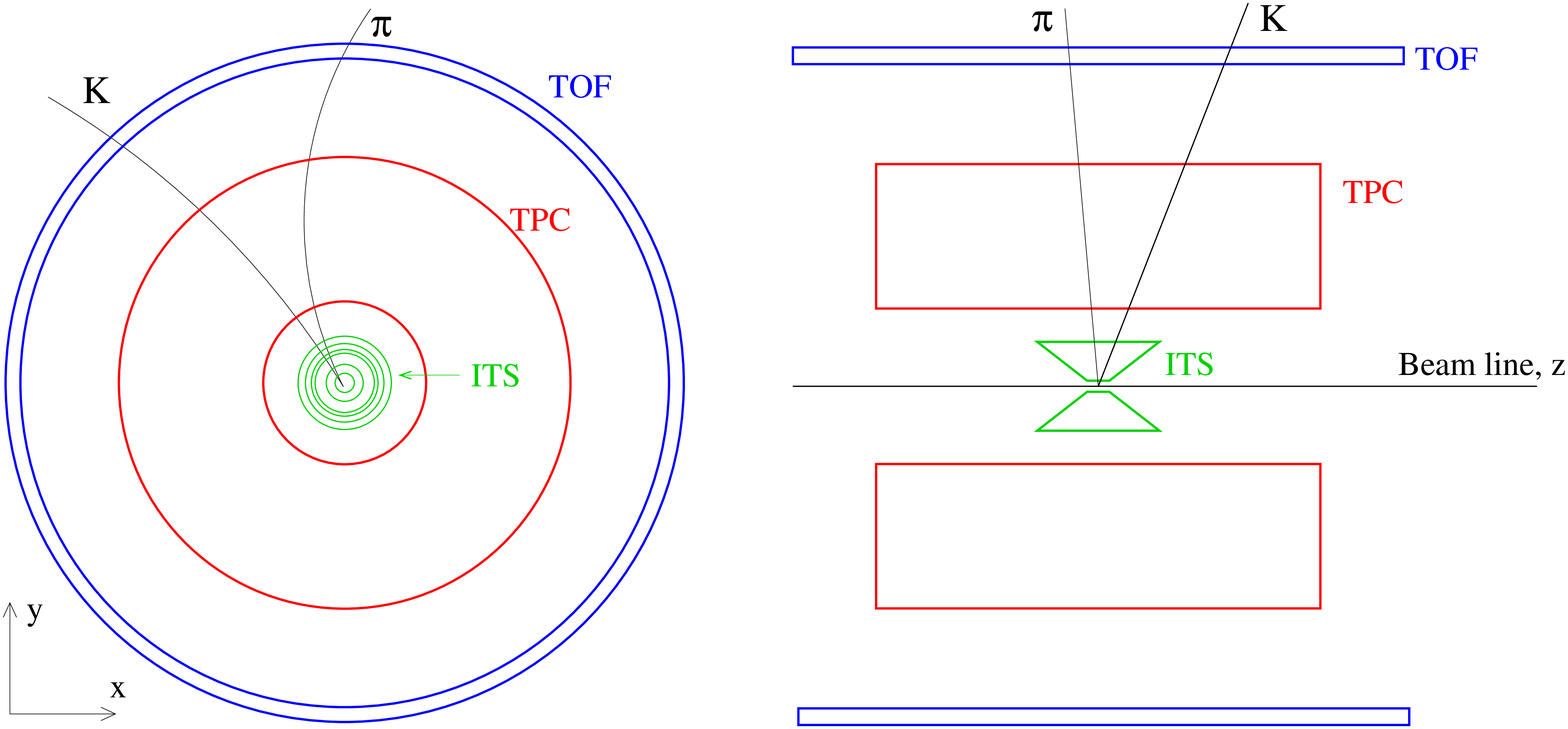}
    \caption{Schematic view of the detectors employed for the reconstruction
             of $\DtoKpi$ decays in ALICE.} 
    \label{fig:D0inALICE}
  \end{center}
\end{figure*}

\begin{figure*}[!t]
  \begin{center}
    \includegraphics[width=.48\textwidth]{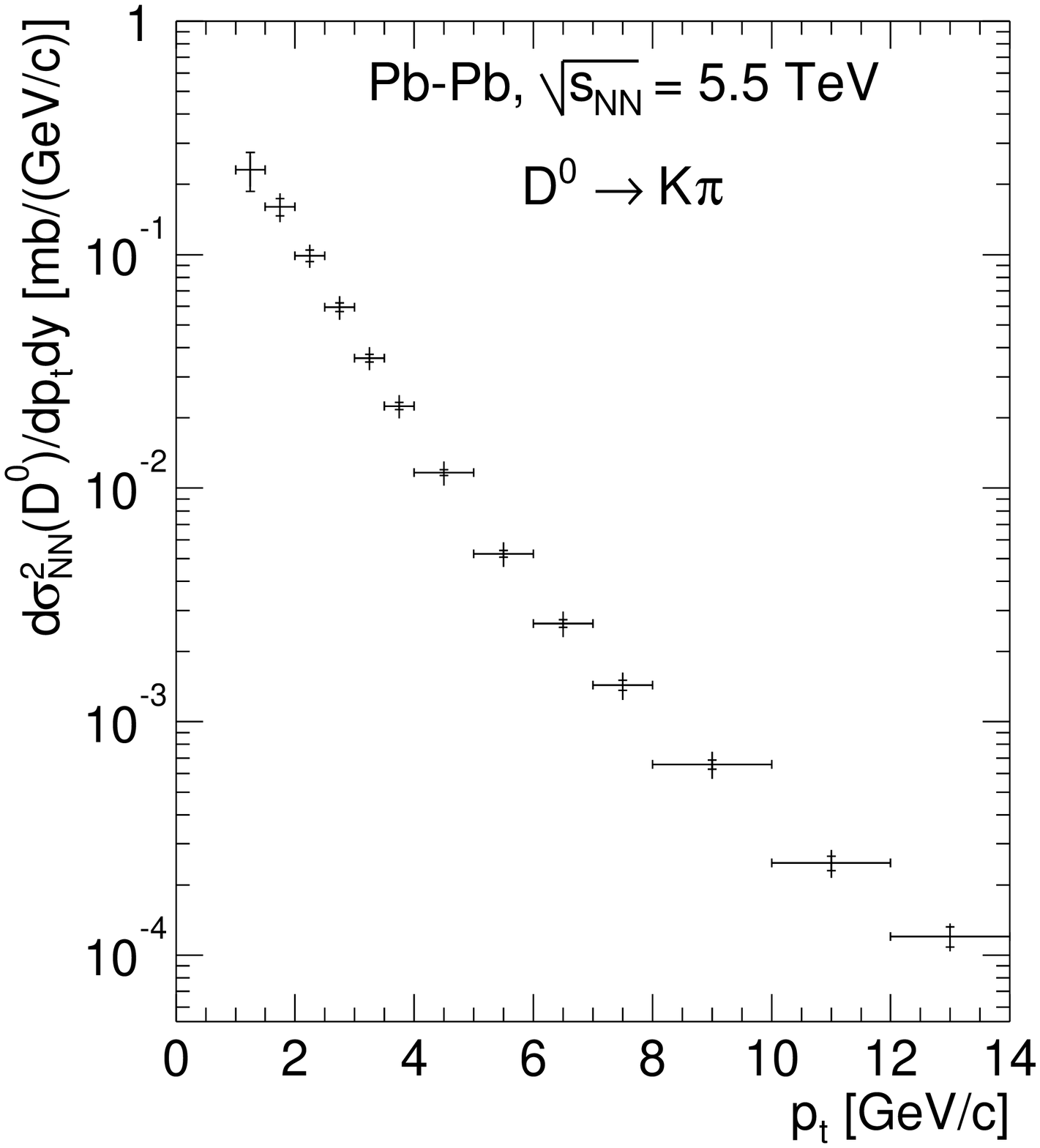}
    \includegraphics[width=.48\textwidth]{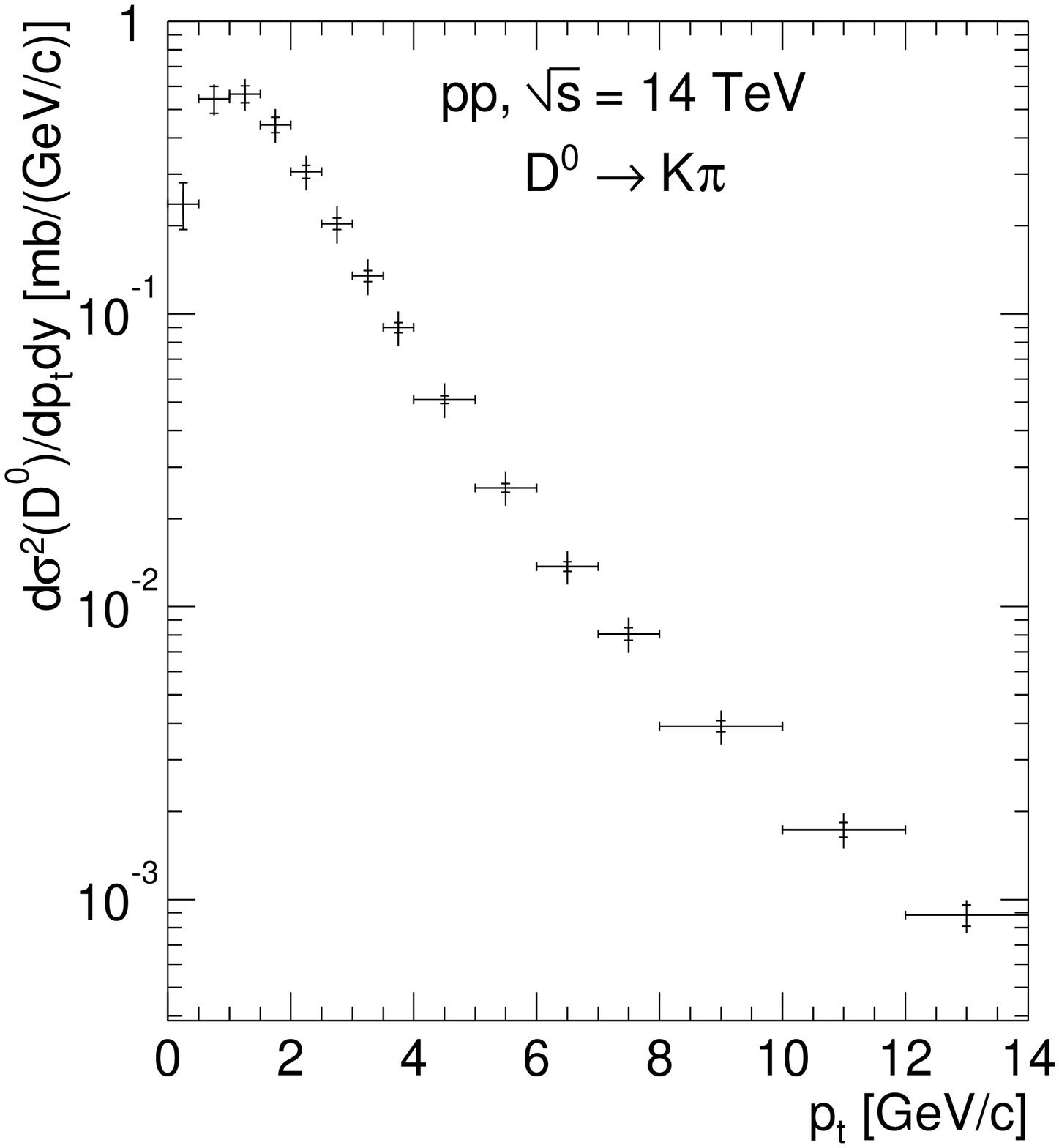}
    \caption{Double-differential cross section per \NN~collision 
             for $\Dz$ production as a 
             function of $\pt$, as it can be measured with $10^7$ central  
             \PbPb~events (left), corresponding to 1-month data-taking, 
             and $10^9$ pp minimum-bias events (right), corresponding to 
             9-months data-taking. 
             Statistical (inner bars) and $\pt$-dependent 
             systematic errors (outer bars) are shown. A normalization error
             of 11\% for \PbPb~and 5\% for pp is not shown.} 
    \label{fig:D0pt}
  \end{center}
\end{figure*}

The transverse momentum distribution of charm mesons produced at central 
rapidity, $|y|<1$, can be directly measured from the exclusive 
reconstruction of $\DtoKpi$ decays 
(and charge conjugates) in the Inner Tracking 
System (ITS), Time Projection Chamber (TPC) and Time Of Flight (TOF) of the 
ALICE barrel ($|\eta|<0.9$). A schema of the employed detectors is reported
in Fig.~\ref{fig:D0inALICE}.  
The displaced vertices of $\Dz$ decays ($c\tau=124~\mum$) can be identified 
in the ITS with silicon pixels, 
that provide a measurement of the track impact parameters to 
the collision vertex with a resolution better than $50~\mum$ for 
$\pt>1~\gev/c$. The low value of the magnetic field (0.4~T) and the 
K/$\pi$ separation in the TOF detector allow to extend 
the measurement of the $\Dz$ production cross section down to almost 
zero transverse momentum. The strategy for this analysis and the
selection cuts to be applied were studied with a realistic and detailed 
simulation of the detector geometry and response, including the main
background sources~\cite{thesis,D0jpg}. 

The expected performance for central \PbPb~($b<3.5~\fm$) 
at $\sqrtsNN=5.5~\tev$ and pp collisions at $\sqrt{s}=14~\tev$
is summarized in Fig.~\ref{fig:D0pt}. The accessible 
$\pt$ range is $1$--$14~\gev/c$
for \PbPb~and $0.5$--$14~\gev/c$ for pp, without assuming dedicated triggers;
triggering on high-$\pt$ tracks or on more specific kinematic and 
topological features 
using the High Level Trigger may allow to further extend these $\pt$ ranges.
The statistical error 
corresponding to 1 month of \PbPb~data-taking and 9 months of pp data-taking 
is better than 15--20\% and the systematic error 
(acceptance and efficiency corrections, subtraction of the feed-down from 
${\rm B}\to \Dz+X$ decays, cross-section normalization, 
centrality selection for \PbPb) is better than 20\%. More details 
are given in Ref.~\cite{thesis}.

\section{Results: $\RAA$ and $\RDh$}
\label{results}

\begin{figure*}[!t]
  \begin{center}
    \includegraphics[width=0.49\textwidth]{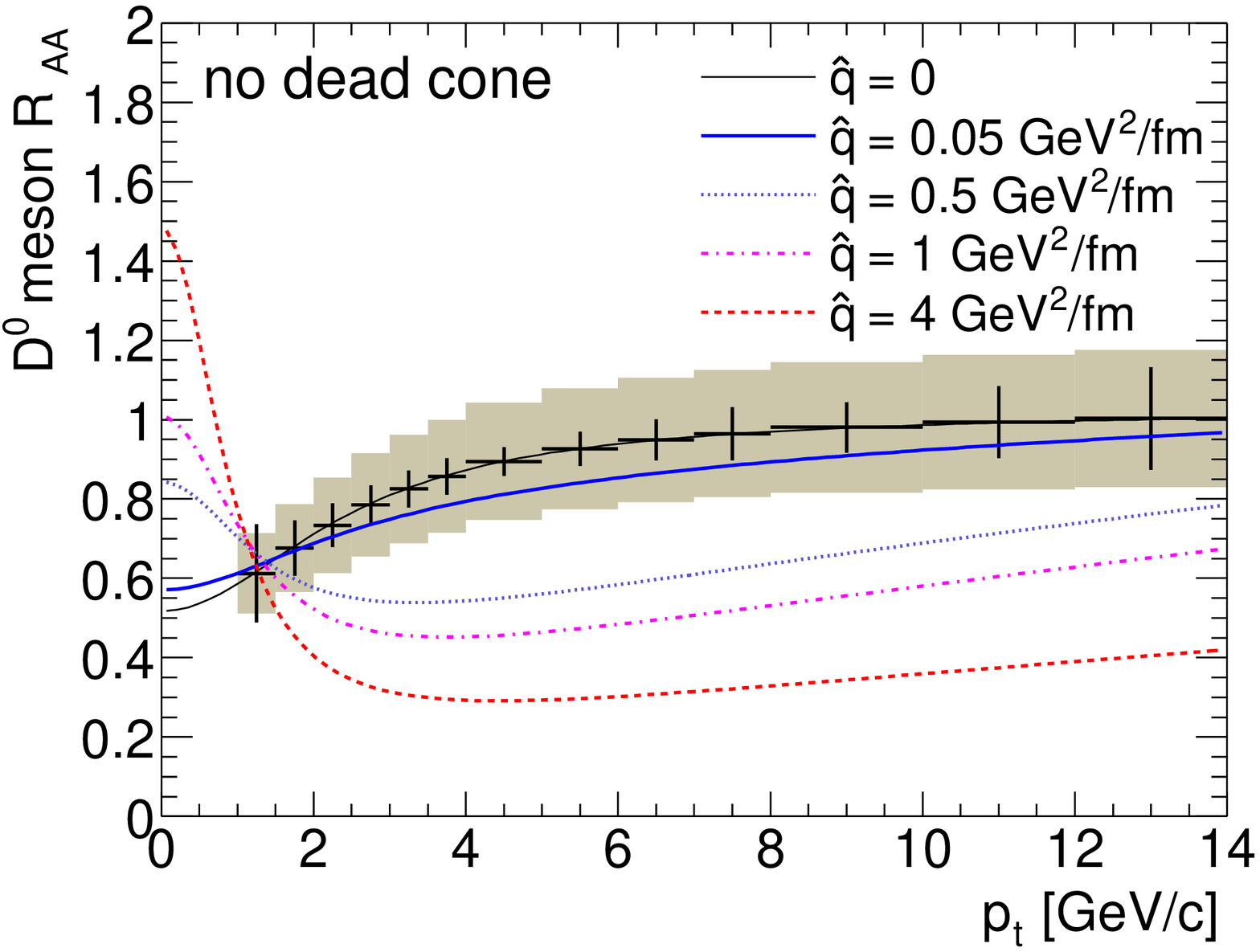}
    \includegraphics[width=0.49\textwidth]{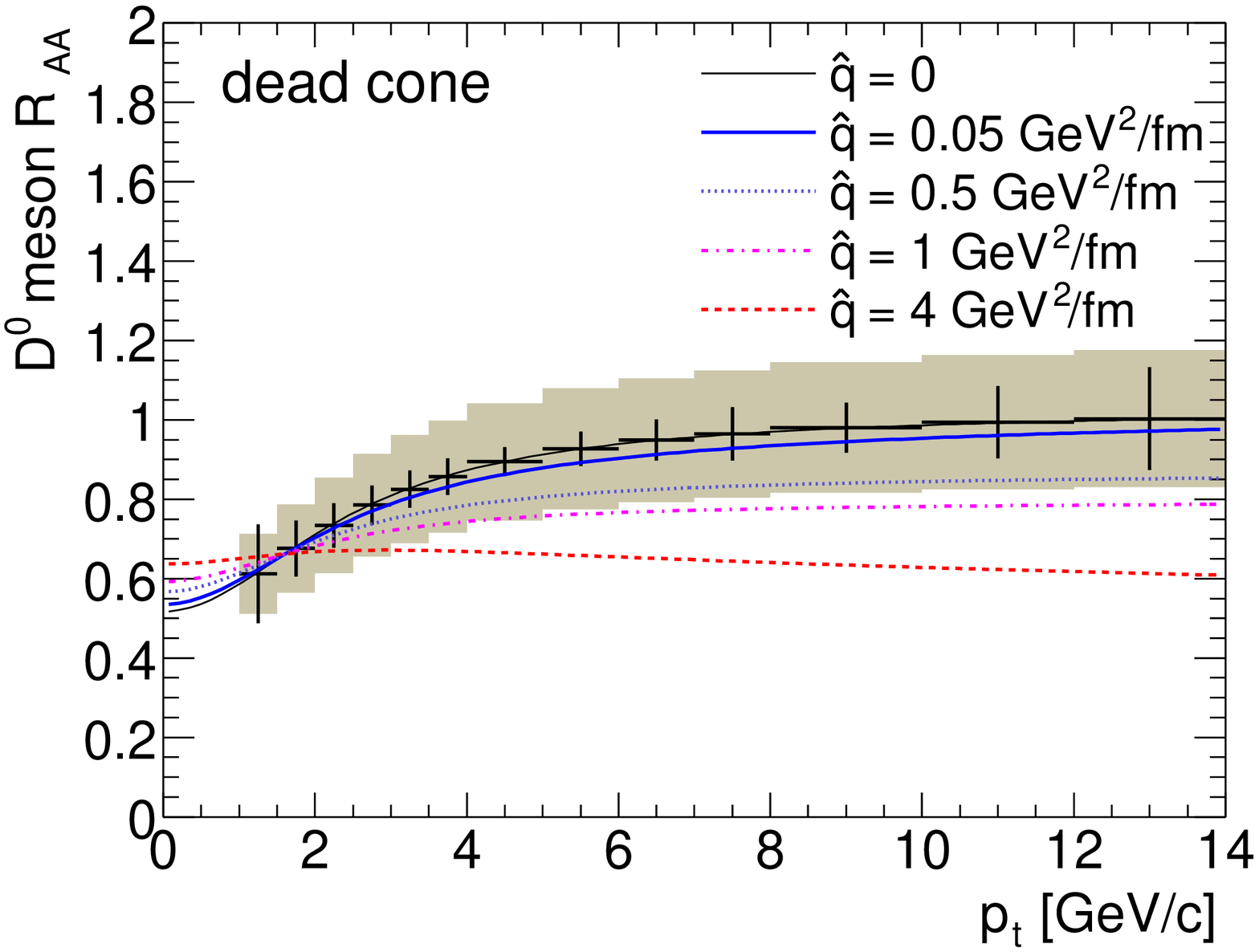}
    \caption{Nuclear modification factor for $\Dz$ mesons with shadowing, 
             intrinsic $k_{\rm t}$ broadening and parton energy loss.
             Left-hand panel: without dead cone correction; right-hand panel: 
             with dead cone correction. Errors corresponding to the curve 
             for $\hat{q}=0$ are shown: bars = statistical, 
             shaded area = systematic.} 
    \label{fig:RAA}
\vglue0.5cm
    \includegraphics[width=0.49\textwidth]{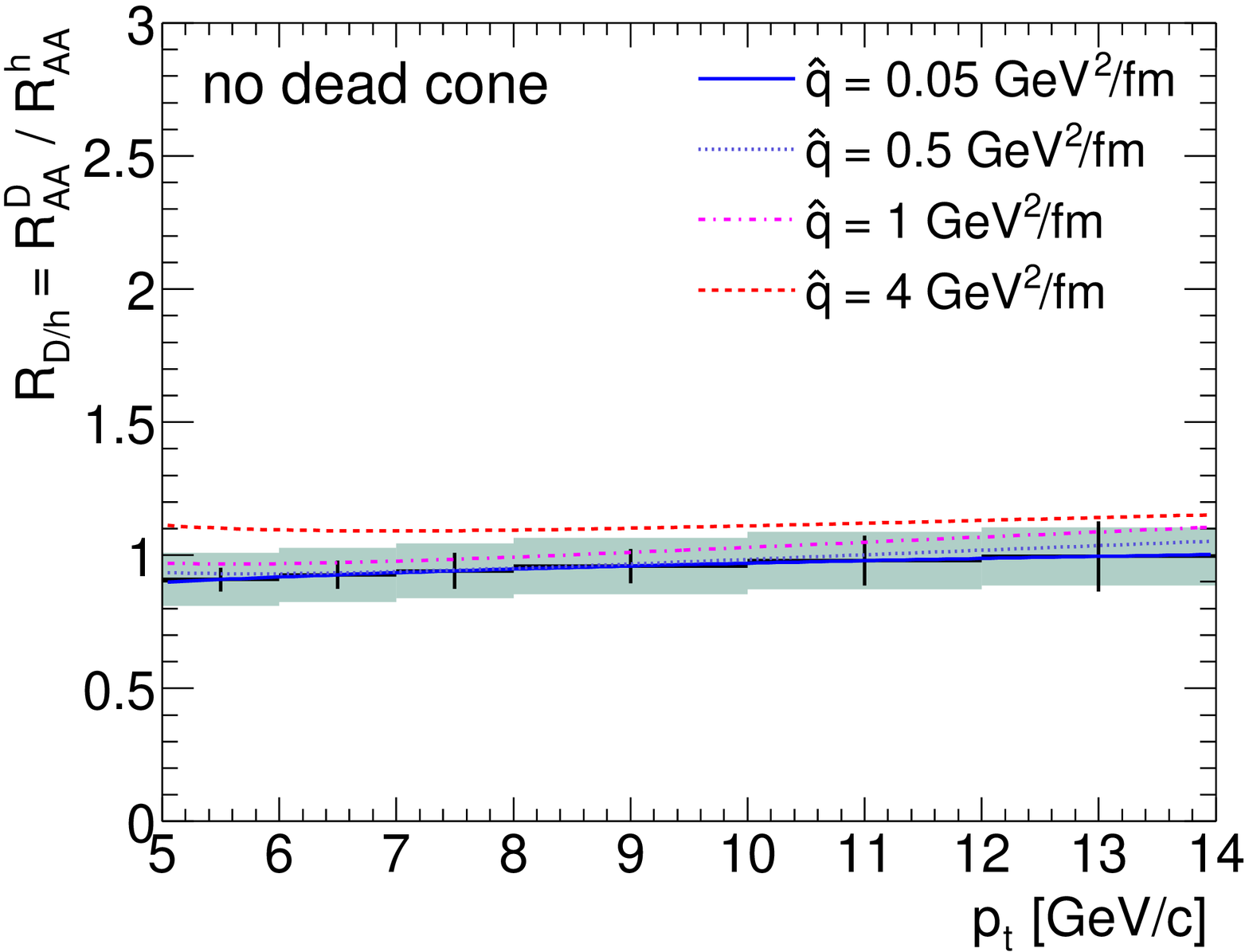}
    \includegraphics[width=0.49\textwidth]{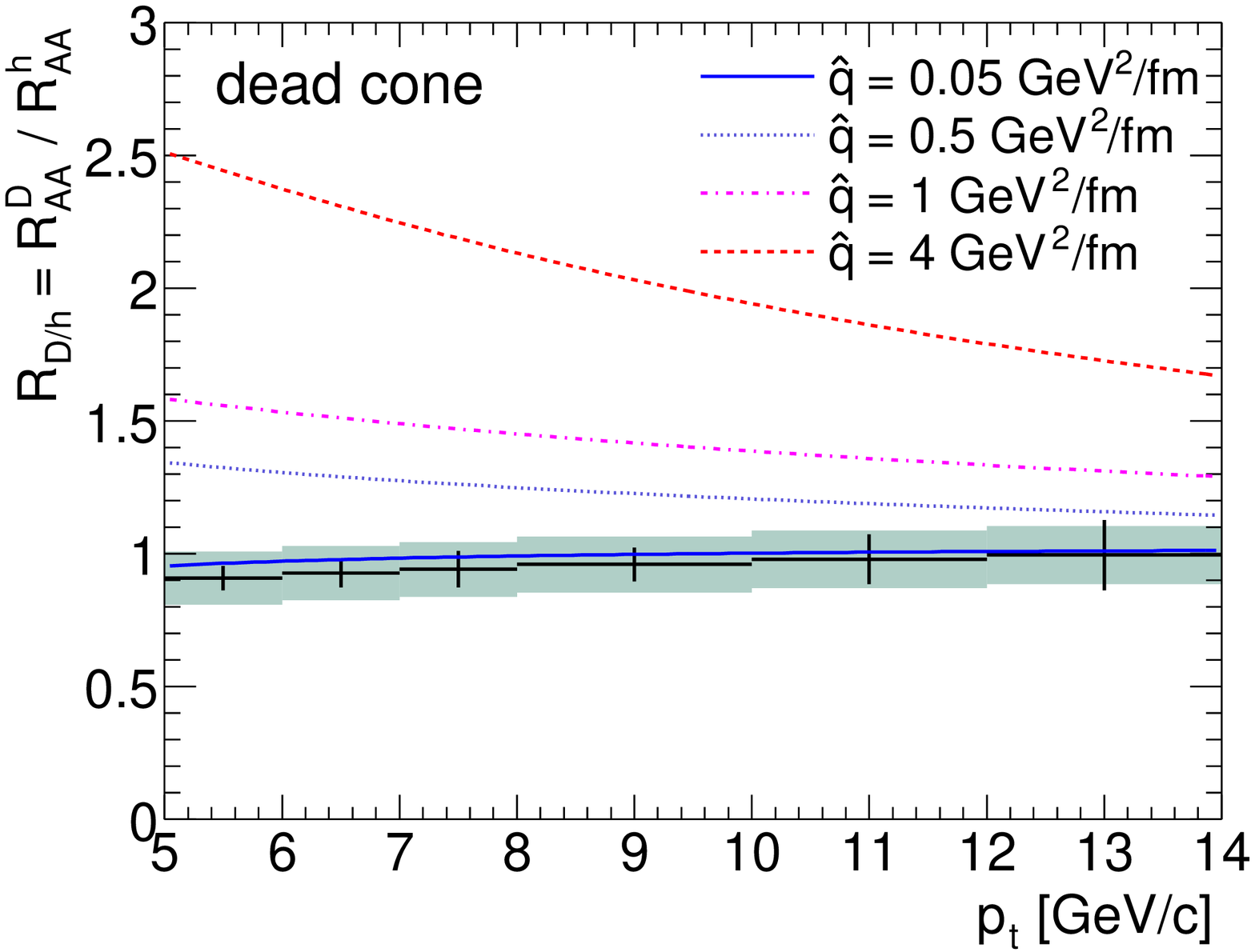}
    \caption{Ratio of the nuclear modification factors for $\Dz$ mesons 
             and for charged hadrons.
             Left-hand panel: without dead cone correction; right-hand panel: 
             with dead cone correction. Errors corresponding to the curve 
             for $\hat{q}=0.05~\gev^2/\fm$ are shown: bars = statistical, 
             shaded area = systematic.} 
    \label{fig:RDh}
  \end{center}
\end{figure*}

The nuclear modification factor (\ref{eq:raa}) for $\Dz$ mesons is reported in 
Fig.~\ref{fig:RAA}. Nuclear shadowing, parton intrinsic 
trans\-verse-momentum broadening and energy loss are included.
The dead-cone effect is not included in the left-hand panel and included in 
right-hand panel. Different values of the transport coefficient are used 
for illustration; we remind that the value expected on the basis of the pion 
quenching observed at RHIC is $\hat{q}=4~\gev^2/\fm$. The reported 
statistical (bars) and systematic (shaded area) errors are obtained combining 
the previously-mentioned errors in \PbPb\, and in pp collisions and 
considering that the contributions due to cross-section normalization, 
feed-down from beauty decays and, partially, acceptance/efficiency corrections
will cancel out in the ratio. An uncertainty of about 5\% introduced in the
extrapolation of the pp results from 14~TeV to 5.5~TeV by pQCD is also 
accounted for (see Ref.~\cite{thesis}). 

The effect of shadowing, clearly visible for $\hat{q}=0$ (no energy loss) 
as a suppression of $\RAA$, is limited to $\pt<6$--$7~\gev/c$ 
(using EKS98~\cite{EKS}).
Above this region, only possible parton energy loss is expected to affect 
the nuclear modification factor of D mesons.

For $\hat{q}=4~\gev^2/\fm$ and no dead cone, we find $\RAA$ 
reduced, with respect to 1, by a factor about 3 and slightly increasing 
with $\pt$, from 0.3 at $6~\gev/c$ to 0.4 at $14~\gev/c$. Even for 
a transport coefficient lower by a factor 4, $\hat{q}=1~\gev^2/\fm$,
$\RAA$ is significantly reduced (0.5--0.6). When the dead-cone effect is
taken into account, the $\RAA$ reduction due to quenching is found to
be lower by about a factor 1.5--2.5, depending on $\hat{q}$ and $\pt$. 
For our reference transport coefficient, $4~\gev^2/\fm$, $\RAA$ with 
dead cone is equal to 0.6 and essentially flat as a function of $\pt$.

We point out that the estimated systematic uncertainty of about 
18\% may prevent from discriminating between a scenario with moderate 
quenching and negligible dead-cone effect (e.g. $\hat{q}=1~\gev^2/\fm$ in the 
left-hand panel of Fig.~\ref{fig:RAA}) and a scenario with large quenching 
but also strong dead-cone effect (e.g. $\hat{q}=4~\gev^2/\fm$ 
in the right-hand panel).

The comparison of the quenching of c-quark-originated mesons and 
massless-parton-originated hadrons will be the best-suited tool to 
disentangle the relative importance of energy-loss and dead-cone effects. 
The D/$charged~hadrons$ ratio $\RDh$, defined as in (\ref{eq:RDh}), is 
presented in Fig.~\ref{fig:RDh} for the range $5<\pt<14~\gev/c$. 
We used $\RAA^h$ calculated as previously described
and $\RAA^{\rm D^0}$, 
without and with dead cone, as reported in Fig.~\ref{fig:RAA}. 
Being essentially a double ratio $\PbPb/\PbPb\times {\rm pp}/{\rm pp}$,
many of the systematic uncertainties on $\RDh$ 
cancel out (centrality selection and, partially, acceptance/efficiency 
corrections and energy extrapolation by pQCD). The residual systematic 
error is estimated to be of about 10--11\%.

We find that, if the dead-cone correction for c quarks is not included, 
$\RDh$ is essentially 1 in the considered $\pt$ range, independently of the 
value of the transport coefficient, i.e. of the magnitude of the energy-loss
effect.
When the dead cone is taken into account, $\RDh$ is enhanced of a factor
strongly dependent on the transport coefficient of the medium:
e.g. 2--2.5 for $\hat{q}=4~\gev^2/\fm$ and 1.5 for $\hat{q}=1~\gev^2/\fm$.
The enhancement is decreasing with $\pt$, as expected (the c-quark mass
becomes negligible).

The $\RDh$ ratio is, therefore, found to be enhanced, with respect to 1,
only by the dead cone and, consequently, it appears as a very 
clean tool to investigate and quantify this effect.

Since hadrons come mainly from gluons while D mesons come from (c) quarks, 
the D/$h$ ratio should, in principle, be enhanced also in absence of 
dead-cone effect, as a consequence of the larger energy loss of gluons with 
respect to quarks. 
Such enhancement is essentially not observed in the obtained $\RDh$ 
because it is `compensated' by the harder fragmentation of charm quarks with 
respect to light quarks and, particularly, gluons. With $z$ the typical 
momentum fraction taken by the hadron in the fragmentation,
$\pt^{\rm hadron}=z\,\pt^{\rm parton}$, and $\Delta E$ the average
energy loss for the parton, $(\pt^{\rm parton})'=\pt^{\rm parton}-\Delta E$,
we have
\begin{equation}
  (\pt^{\rm hadron})'=\pt^{\rm hadron} - z\,\Delta E,
\end{equation}
meaning that the energy loss observed in the nuclear modification factor is, 
indeed, $z\,\Delta E$. We have, thus, to compare 
$z_{\rm c\to D}\,\Delta E_{\rm c}$
to $z_{\rm gluon \to hadron}\,\Delta E_{\rm gluon}$. With 
$z_{\rm gluon \to hadron}\approx 0.4$, $z_{\rm c\to D}\approx 0.8$ 
for $\pt^{{\rm D},h}>5~\gev/c$ 
and $\Delta E_{\rm c}=\Delta E_{\rm gluon}/2.25$ (without dead cone),
we obtain 
\begin{equation}
  z_{\rm c\to D}\,\Delta E_{\rm c}\approx 0.9\,z_{\rm gluon\to hadron}\,\Delta E_{\rm gluon}.
\end{equation} 
This simple estimate confirms that the quenching for D mesons is 
almost the same as for (non-charm) hadrons, if the dead-cone 
effect is not considered. 
                      
The errors reported in Fig.~\ref{fig:RDh} show that ALICE is expected to have
good capabilities for the study of $\RDh$: in the range $5<\pt<10~\gev/c$
the enhancement due to the dead cone is an effect of more than $3~\sigma$
for $\hat{q}>1~\gev^2/\fm$.
The comparison of the values for the 
transport coefficient extracted from the nuclear modification factor of 
charged hadrons and, independently, from the 
D/$charged~hadrons$ ratio can provide an important test for the coherence of 
our understanding of the energy loss of hard probes propagating in the 
dense QCD medium formed in \PbPb~collisions at the LHC.

\subsection*{Acknowledgements}

The present study was carried out within the ALICE Collaboration and using 
the software framework and analysis algorithms developed by the off-line
project. I acknowledge the ALICE off-line group for support and useful 
discussions. I am grateful to the following members of the ALICE Collaboration
for comments and suggestions: F.~Antinori, N.~Carrer, A.~Morsch, G.~Paic, 
E.~Quercigh and
K.~$\check{\mathrm{S}}$afa$\check{\mathrm{r}}$\'\i k.
I enjoyed many stimulating 
discussions with N.~Armesto, K.J.~Eskola, C.A.~Salgado and U.A.~Wiedemann
on the phenomenology of parton energy loss.


\end{document}